\documentclass[twocolumn,trackchanges]{aastex631}

\newcommand{\rb}{$R_b$}
\newcommand{\alpham}{[$\alpha$/M]}

\begin{document}

\title{Stellar Birth Radii in the LMC: Insights into Chemodynamics, Radial Migration, and Star Formation Across the Disk}

\newcommand{\MSU}{Montana State University, P.O. Box 173840, Bozeman, MT 59717-3840, USA}
\newcommand{\osu}{Department of Astronomy, The Ohio State University, Columbus, 140 W 18th Ave, OH 43210, USA}
\newcommand{\ccapp}{Center for Cosmology and Astroparticle Physics (CCAPP), The Ohio State University, 191 W. Woodruff Ave., Columbus, OH 43210, USA}

\author[0000-0003-4769-3273]{Yuxi(Lucy) Lu}
\affiliation{\osu}
\affiliation{\ccapp}

\author[0000-0001-6013-9125]{Bethany Garver}
\affiliation{\MSU}

\author[0000-0002-1793-3689]{David L. Nidever}
\affiliation{\MSU}

\author[0000-0002-6553-7082]{Joshua T. Povick}
\affiliation{\MSU}


\author[0000-0001-7107-1744]{Nicolás Garavito-Camargo}
\affiliation{Steward Observatory, University of Arizona, 933 North Cherry Avenue, Tucson, AZ 85721, USA}
\affiliation{Department of Astronomy, University of Maryland, College Park, College Park, MD 20742, USA}

\author[0000-0002-6797-696X]{Maria-Rosa L. Cioni}
\affiliation{Leibniz-Institut f\"{u}r Astrophysik Potsdam, An der Sternwarte 16, D-14482 Potsdam, Germany}

\author[0000-0003-1856-2151]{Danny Horta}
\affiliation{Institute for Astronomy, University of Edinburgh, Royal Observatory, Blackford Hill, Edinburgh, EH9 3HJ, UK}



\begin{abstract}
The LMC and SMC are interacting dwarf galaxies that offer a valuable testbed for studying the effects of galactic mergers. 
We investigate the chemodynamic history of the LMC in the context of its interaction with the SMC by inferring stellar birth radii, first validated on a hydrodynamical simulation tailored to reproduce their interaction history. 
Using inferred birth radii and stellar ages, we identify signatures of dynamical and chemical evolution across the LMC disk.
We find that the LMC’s metallicity gradient steepened around 5, 3, and 1 Gyr ago, coinciding with enhanced star formation (SF) episodes. 
These events exhibit distinct spatial patterns --- initially concentrated in the inner disk at 5 Gyr, expanding outward by 3 Gyr, and becoming widespread with renewed central activity at 1 Gyr --- likely reflecting changes in spin alignment between the interacting disks if the enhancements of SF tracks the pericenter passages of the SMC to the LMC. 
The inferred radial migration strength of the LMC shows notable enhancements at 0.5, 2, and 5 Gyr.
The most $\alpha$-enriched stars form 2–3 Gyr ago at birth radii of 2–4 kpc, the only epoch when star formation is broadly distributed across the disk. 
Finally, unlike the Milky Way, the LMC lacks a clear \alpham–[Fe/H] bimodality.
This is likely due to its more centrally concentrated star formation during these periods, compared to the MW’s more extended outer-disk star formation enhancements.
These findings place strong constraints on the LMC's assembly history and its interaction with the SMC.

\end{abstract}



\section{Introduction} \label{sec:intro}
Radial migration is the process by which stars undergo permanent changes in angular momentum due to dynamical interactions, often at co-rotation resonances of various galactic structures.
These interactions can arise from transient spiral arms \citep[e.g.,][]{Sellwood2002, Roskar2008}, overlapping resonances between the bar and spiral structure \citep[e.g.,][]{Minchev2010}, external perturbations \citep[e.g.,][]{Quillen2009}, or a slowing bar \citep[e.g.,][]{Zhang2025}.
Stars moving away from their birth locations means one cannot and should not equate an age-resolved evolution of a galaxy with its temporal evolution \citep[e.g.,][]{Minchev2013, Frankel2019, Sharma2020, Lu2022, Ratcliffe2023}.
One such common misconception is using ``look-back time'' and ``age'' interchangeably.
Gradient evolution in age is typically shallower and exhibits more scatter around the relation than that in look-back time, as stars at a given age have had time to migrate across the galaxy (increase in scatter).
They also typically move outwards as a population (flattening the gradient), due to the exponential surface density profile in galaxies \citep[e.g.,][]{Minchev2013}.
As a result, in order to infer the true formation history of a galaxy, it is important to take into account the effects of radial migration.

Extensive efforts have been devoted to identifying the effects and signatures of radial migration in the Milky Way (MW).
For example, studies have focused on understanding the metallicity distribution functions and the age-metallicity relation \citep[e.g.,][]{Sellwood2002, Loebman2016, Lu2022, Lian2022}, the formation of the high- and low-$\alpha$ disks \citep[e.g.,][]{Sharma2020, Sahlholdt2022, Lu2024}, the temporal evolution of the metallicity gradient \citep[e.g.,][]{Minchev2018, Lu2024, Ratcliffe2023}, and many more \citep[e.g.,][]{Frankel2020}.
Despite extensive efforts in the MW, the effects of radial migration in external galaxies have received comparatively little attention.
Recently, studies have shown that inferring the spatially resolved star formation history for disk galaxies without taking into account radial migration can introduce large uncertainties \citep{Minchev2025, Bernaldez2025, Ratcliffe2025}.

To begin addressing this gap, we start with the closest galaxy pairs to us --- the MCs system.
The LMC and the SMC have photometric observations for individual stars down to the main-sequence \citep[e.g., VMC, SMASH,][]{Cross2012, Cioni2016, Nidever2017, Nidever2021}, providing us with detailed star formation history (SFH) of the MCs \citep{Rubele2018, RuizLara2020, Mazzi2021, Massana2022, Saroon2025}.
Moreover, detailed abundances are also available \citep{Nidever2020} for a subset of the upper red giant branch stars (RGB) through the APOGEE--2S \citep{Wilson2019}.
These observations of individual stars provide us with the exact data set we need to obtain birth radii (\rb, the galactic-centric radius of where the stars are born).
Combining age, \rb, element abundances, and the SFH, we can begin to uncover the true chemodynamic history of the LMC and assess the role of major mergers in shaping galaxy evolution.

In \cite{Lu2024_lmc}, we explored the possibility of obtaining \rb\ for external galaxies using the method described in \cite{Lu2024}, which is a method improved upon \cite{Minchev2018}, using a set of updated assumptions \citep{Lu2022_rblim}.
Using the Numerical Investigation of a Hundred Astronomical Objects \citep[NIHAO;][]{Wang2015} cosmological simulations, we determined it is important to consider the effects of radial migration in the LMC when understanding its formation history, and birth radii can be inferred with $\ sim$25\% uncertainty. 

In this paper, we test the method described in \cite{Lu2024} on a hydrodynamical simulation of the LMC \citep{Garver2026} that is set up to reproduce the formation and interaction history of the LMC and SMC.
The description of the simulation is in Section~\ref{subsec:sim}.
After successfully inferring \rb\ from the simulation (Section~\ref{subsec:recover_sim}), we apply this method to the LMC observations (Section~\ref{subsec:obs}) to infer \rb\ and the temporal evolution of the metallicity gradient (Section~\ref{subsec:recover_obs}). 
In Section~\ref{sec:results}, we look at the radial migration strength, preferred star formation radii across time, and the chemodynamic history of the LMC in \alpham-[Fe/H] space.
Finally, in Section~\ref{sec:twoinfall}, we discuss how insights from the LMC–SMC merger inform our understanding of the two-infall model for the MW.

\section{Data \& Methods} \label{sec:datamethod}
\subsection{Simulation Data} \label{subsec:sim}

The orbit of the MCs is taken from the \citet{Garver2025} suite of $\sim$8,000 N-body interaction simulations.  This produces closer encounters $\sim$2 Gyr and $\sim$150 Myr ago.
We use the smooth particle hydrodynamics+N-body tree-code \small{GASOLINE} \citep{Wadsley2004,Wadsley2017} to simulate the MCs and MW.  The LMC and SMC are simulated in isolation for 6 Gyr and are then combined with the MW and evolved for another 2.5 Gyr.
The simulated LMC begins with a gas disk of mass $7.18\times 10^9 M_{\odot}$ and a dark matter halo of mass $1.81 \times 10^{11} M_\odot$. It is evolved in isolation for 6 Gyr during which time some of the gas forms into stars. Then it is combined with an SMC and MW model, and the combined simulation is evolved for a further 2.5 Gyr.
\small{GASOLINE} tracks the [Fe/H] and [O/Fe] of the star and gas particles, as well as the birth times and locations of the stars. The simulation uses 20\% feedback, which means that 20\% of the $10^{51}$ erg of energy from SNe is injected as thermal energy into the interstellar medium.



\subsection{Observational Data} \label{subsec:obs}

Stellar ages were derived following the method in \cite{Povick2024} using observed APOGEE spectroscopic parameters (Teff, $\log{g}$, {[Fe/H]}, and {[$\alpha$/Fe]}) and multi-band photometry from Gaia \citep{gaia, Gaiadr3} and 2MASS \citep{2mass} with PARSEC isochrones \citep{parsec, parsec2, parsec3}. 
Distances for the stars were found using the red clump model from SMASH \citep{Choi2018a, Choi2018b} and assuming that the RGB stars lie close to the disk plane.
For each star, model photometry and $\log g$ were interpolated from the isochrones as functions of Teff, {[Fe/H]}, {[$\alpha$/Fe]}, and a trial age. 
Since the isochrones are based on scaled-solar models, the {[Fe/H]} are adjusted with the measured {[$\alpha$/Fe]} following \cite{Salaris1993} with updated coefficients from \cite{Povick2024}.
The trial age was then adjusted until the model photometry and $\log{g}$ best matched the observed values. Validation with the APOKASC asteroseismic sample (with accurate masses and ages) indicates that the age uncertainties are $\sim$20\% \citep[see Figure 16 in][]{Povick2024}.
The selection function is calculated following the method described in Section 9 in \cite[][]{Povick2024}.

\begin{figure*}[ht!]
    \centering
    \includegraphics[width=\textwidth]{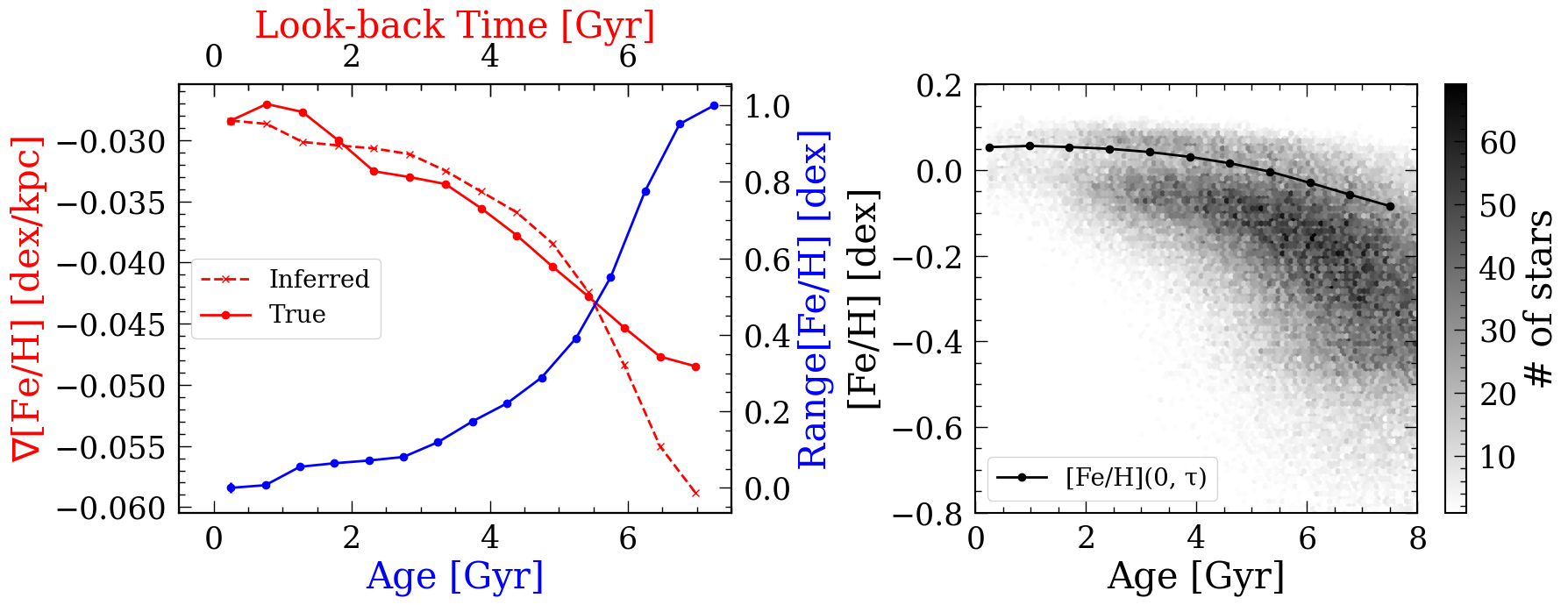}
    \caption{Key plots for inferring the metallicity evolution of the simulated LMC galaxy. 
    Left: The blue line shows the range of metallicity, Range[Fe/H], as a function of age, using a bin width of 0.5 Gyr. 
    The range is measured using the difference between the 95$^{\rm th}$-percentile and the 10$^{\rm th}$-percentile of [Fe/H] in mono-age populations to account for more stars in the inner galaxy compared to the outer. 
    The red solid line shows the true metallicity gradient, $\nabla$[Fe/H], as a function of look-back time, measured using the true \rb\ and [Fe/H] in each mono-age population.
    The red dashed line is the inferred metallicity gradient using Equation~\ref{eq1}.
    We can reproduce some of the key features, such as the steepening, in the metallicity gradient, and the median absolute deviation from the true metallicity gradient is 6\%, or 0.002 dex/kpc.
    Right: The evolution of metallicity at the galactic center is shown in the black solid line. 
    The background shows the histogram of the age-metallicity relation for this simulated galaxy.
    Again, the most metal-enriched stars at every age bin should represent the metallicity of the LMC center at that look-back time, assuming a negative metallicity gradient is present at all ages after the disk has started to form.}
    \label{fig:1}
\end{figure*}

\subsection{Recovering Birth Radii for LMC Simulation} \label{subsec:recover_sim}
Since \rb\ inference can only be applied to stars formed in the galactic disk, we selected stars in the simulation that are born within 15 kpc of the galactic center with birth vertical height, $|z_{\rm bir}| <$ 1 kpc.
Changing to select $|z_{\rm bir}| <$ 0.5 kpc does not alter our result significantly.
We then scale the metallicity so that the standard deviation and median match those of the observations (median: $-$0.77 dex, std: 0.23 dex).
To mimic the observational uncertainty, we perturbed the metallicity and age in the simulation by their typical observational uncertainty (0.015 dex for metallicity and 20\% for age).

To infer \rb, we first applied the procedure described in \cite{Lu2024} to measure the range in metallicity in mono-age populations.
The method stated that the metallicity gradient as a function of look-back time can be directly inferred using the range in metallicity as a function of age with a linear relation.
Since fewer stars are formed in the outer disk, which should be the most metal-poor if we assume a negative metallicity gradient due to inside-out formation, we selected 10$^{\rm th}$-percentile for the lower limit instead of 5$^{\rm th}$-percentile when calculating the range of metallicity (Range[Fe/H]). 
We inferred the metallicity gradient ($\mathrm{\nabla [Fe/H]}$) as a function of look-back time ($\tau$) using Equation (1) in \cite{Lu2024}:
 \begin{equation}\label{eq1}
    \mathrm{\nabla [Fe/H]}(\tau) = a\ \mathrm{Range{[Fe/H]}(age)}+\mathrm{\nabla [Fe/H]}(0),
\end{equation}
Where $\mathrm{\nabla [Fe/H]}(0)=-$0.028 dex/kpc is the metallicity gradient at redshift 0, inferred using the youngest stars, and $a$ is the scaling factor, which we used $-$0.031 kpc$^{-1}$ to ensure the disk is formed inside-out (oldest stars have the smallest \rb).
The scaling factor 
The range, the inferred gradient evolution, and the true gradient evolution for the simulation are shown in the left plot of Figure~\ref{fig:1}.
For better understanding, the range in metallicity (Range[Fe/H]) provides the shape of the inferred metallicity gradient, the current-day metallicity gradient, $\nabla$[Fe/H](0), provides the anchor point, and the scaling factor, $a$, constrains the range, or the steepest point, for the metallicity gradient.

Birth radii are then inferred with age and metallicity measurements using Equation (3) in \cite{Lu2024}:
\begin{equation}\label{eq2}
    \mathrm{R_b(age, [Fe/H])} = \frac{\mathrm{[Fe/H]} - \mathrm{[Fe/H]}(0,\tau)}{\nabla\mathrm{[Fe/H](\tau)}}.
\end{equation}
Where $\mathrm{[Fe/H]}(0,\tau)$ is inferred by first taking the 95$^{\rm th}$-percentile of the metallicity in mono-age populations (as shown in the right plot of Figure~\ref{fig:1}).
This is because, assuming the metallicity gradient is always negative (naturally arising from inside-out formation), the metallicity of the center of the galaxy at a given look-back time will always contain the most metal-enriched stars of the same age.
We then shift $\mathrm{[Fe/H]}(0,\tau)$ so that stars $<$ 1 Gyr have $R-R_b$ peaks close to 0, as these young stars should not have much time to migrate.
The shifted value is determined to be -0.04 dex.

\begin{figure}[ht!]
    \centering
    \includegraphics[width=\linewidth]{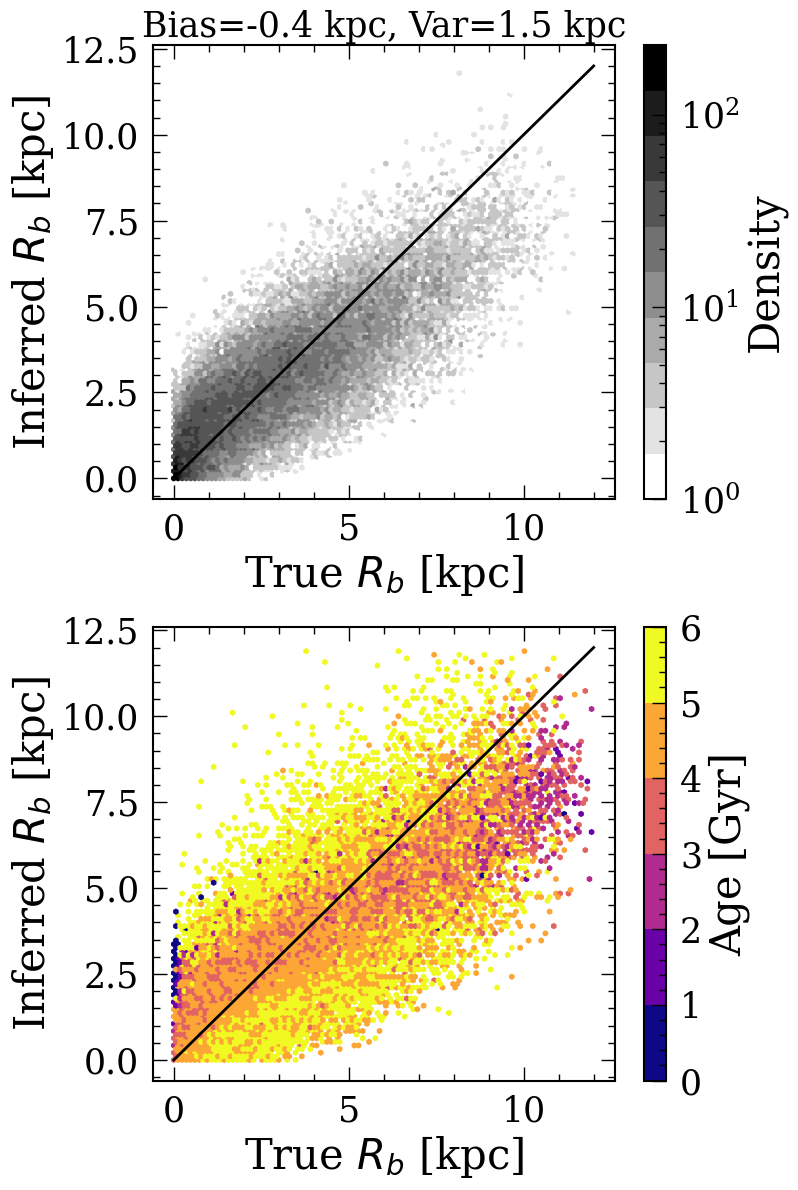}
    \caption{Histogram of the true birth radii from the simulated LMC vs the inferred birth radii using the method described in Section~\ref{subsec:recover_sim}, colored by the number of stars (top) and the median age (bottom).
    We are able to successfully infer birth radii with a bias of -0.6 kpc and a variance of 1.5 kpc for stars $<$ 6 Gyr.}
    \label{fig:2}
\end{figure}

Figure~\ref{fig:2} shows the inferred birth radii using the method described in this section compared to the truth for the simulation.
We are able to infer birth radii with a variance of 1.5 kpc and a bias of -0.4 kpc for stars $<$ 6 Gyr.
The bias could affect the absolute values of our results, but since the birth radii in this paper are primarily used to rank the locations where stars are born, the impact should be small as long as the ranking is preserved, i.e., stars formed in the outer disk have larger inferred \rb\ compared to those formed in the inner galaxy, the bias should not significantly affect our conclusions regarding star-forming regions as a function of look-back time or the inferred migration strength.

\begin{figure*}[ht!]
    \centering
    \includegraphics[width=0.32\linewidth]{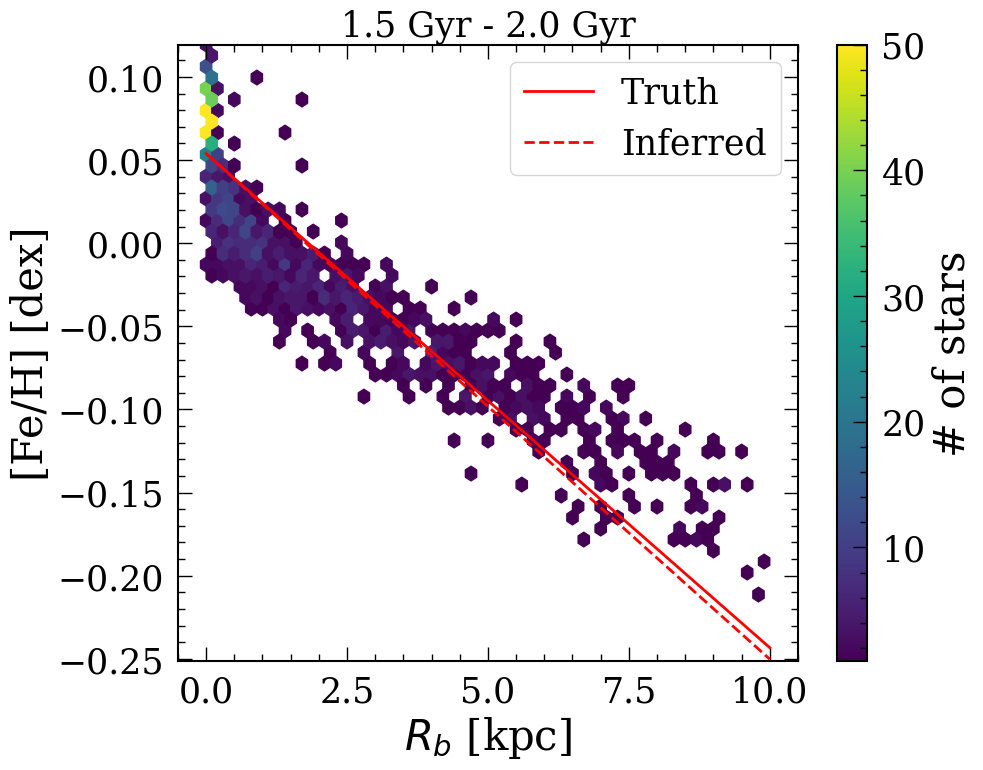}
    \includegraphics[width=0.32\linewidth]{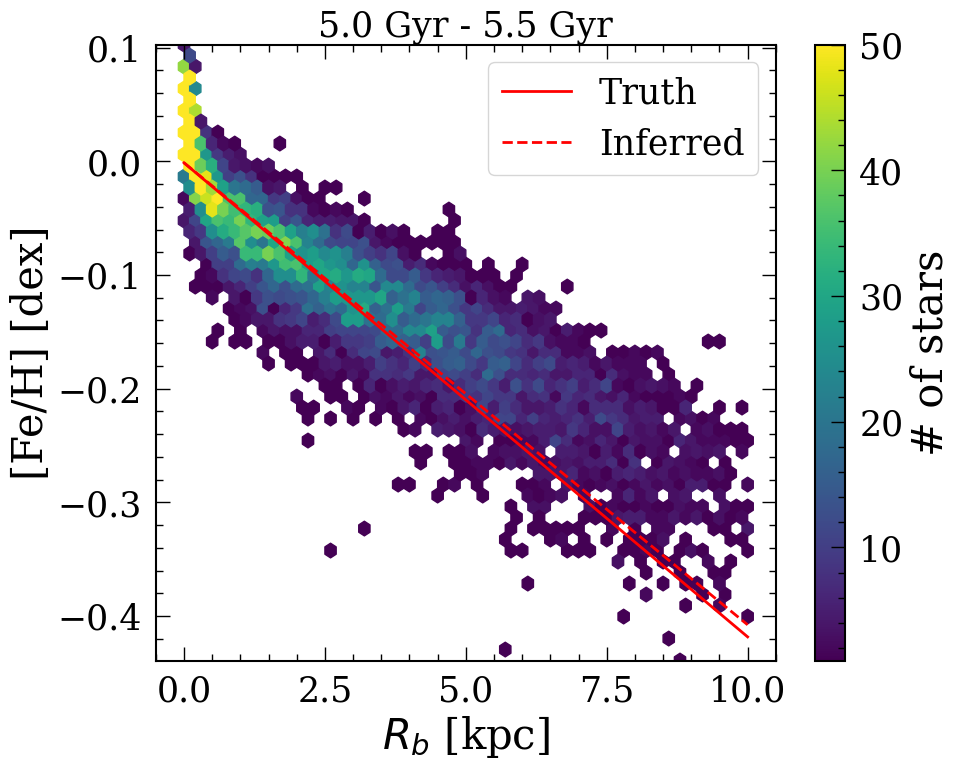}
    \includegraphics[width=0.32\linewidth]{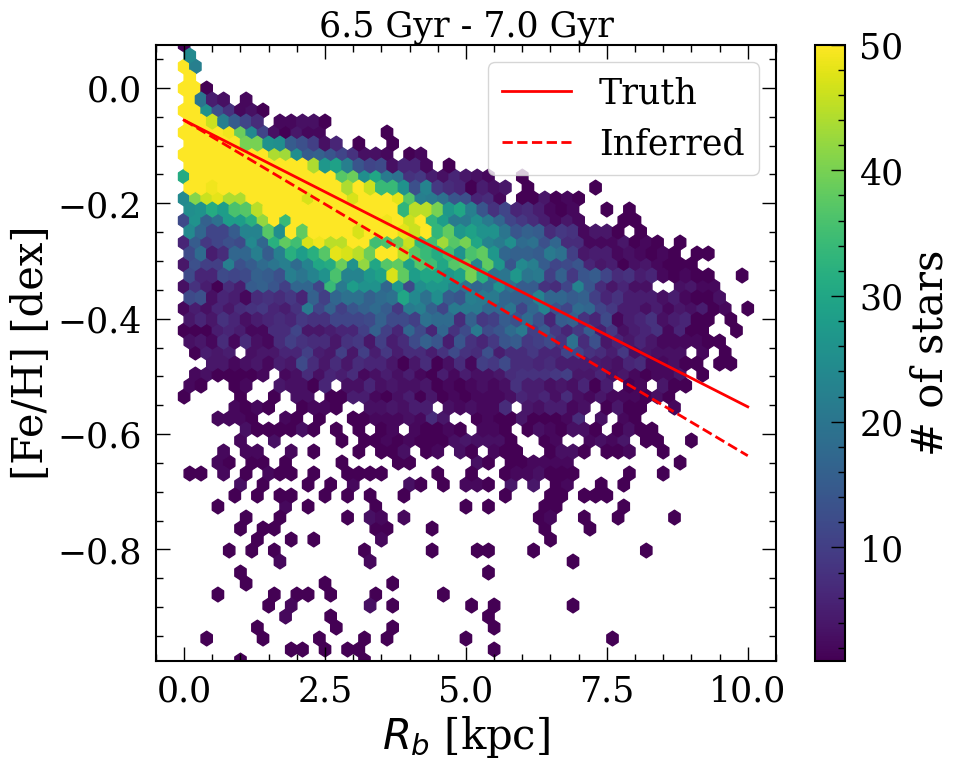}
    \caption{Metallicity-\rb\ relation for stars in the LMC simulation during interaction phase between the LMC and SMC (1.5-2 Gyr; left), isolation phase (5-5.5 Gyr; middle), and settling phase (6.5-7 Gyr; right).
    The solid lines show the true metallicity gradient, and the dashed line shows the inference from Range[Fe/H]. 
    The linear relation does fit the outer galaxy well, as the Metallicity-\rb\ relations are not exactly linear, especially in the inner galaxy. 
    This causes the bias for inferring \rb\ for the outer galaxy (see Figure~\ref{fig:2}).}
    \label{fig:inferslopes}
\end{figure*}

Figure~\ref{fig:inferslopes} shows the [Fe/H]-\rb\ distribution (background density plot), the true gradient (red, solid lines), and the inferred gradient (red, dashed lines) for three different look-back time bins.
To infer birth radii, we assume all stars follow a perfect relation shown as the dashed line.
The fits do not perform well for stars in the outer galaxy, as the gradient is not exactly linear, and most stars are born in the inner galaxy.
This could explain the bias seen for stars in the outer disk (Figure~\ref{fig:2}).
One other source of uncertainty comes from the fact that there is scatter around the relation.
For stars $>$ 6 Gyr, the stellar disk is still settling from the initial condition, and since this method can only be applied to stars after the stellar disk has mostly formed \citep{Lu2022_rblim}, it is not surprising we are not able to reproduce the true metallicity gradient while the disk is still settling and the scatter around the metallicity gradient is larger (see Figure~\ref{fig:inferslopes} right panel).

\begin{figure*}[ht!]
    \centering
    \includegraphics[height=0.25\linewidth]{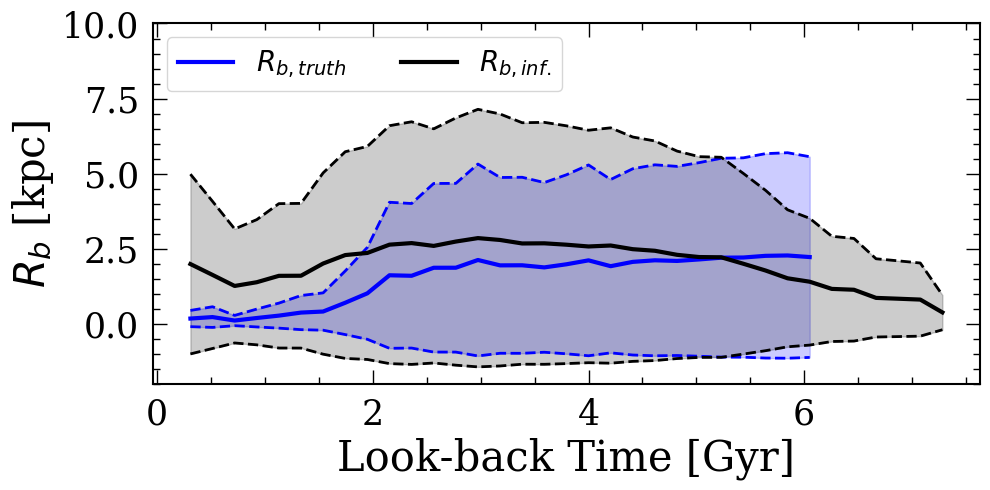}
    \includegraphics[height=0.25\linewidth]{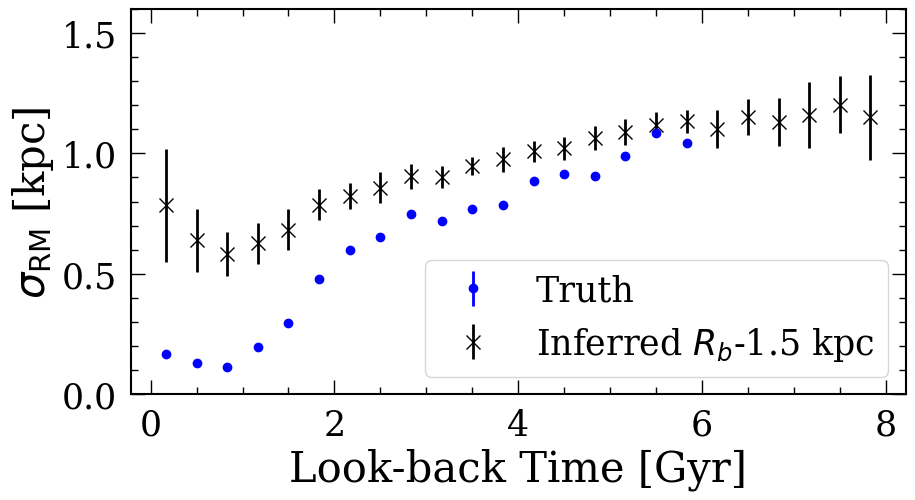}
    \caption{Left: The median and the 1.5*MAD of the birth radii distribution as a function of age using the true \rb\ ($R_{b, truth}$) and age shown in blue, and the inferred \rb\ ($R_{b, inf.}$) and true age convolved with a 20\% uncertainty in black. 
    Right: The true migration strength (blue points) compared to the migration strength measured using inferred \rb\ (black crosses with errorbars) for the simulation.
    1.5 kpc is subtracted from the strength using inferred \rb\ to take into account the variance in our inference (see Figure~\ref{fig:2}).
    The uncertainties are measured with bootstrapping \rb\ with a width of 1.5 kpc, and age with a width of 20\% of its value.
    Both relations using inferred \rb\ extend beyond 6 Gyr due to the imposed age uncertainty.
    We are able to reproduce the general shape for both the \rb\ distributions and the radial migration strength.}
    \label{fig:2.5}
\end{figure*}

Finally, Figure~\ref{fig:2.5} compares the \rb\ distribution as a function of age and the radial migration strength in the simulation for stars $<$ 6 Gyr.
The left panel presents the median and 1.5*MAD of \rb\ in each age bin using the truth \rb\ ($R_{b, truth}$; blue) and the inferred \rb\ ($R_{b, inf.}$; black).
Ages include a 20\% uncertainty convolved to mimic observation data.
The overall shape of the median and dispersion is reproduced down to $\sim$1 Gyr.
This is not surprising as we are only able to infer birth radii to the accuracy of 1.5 kpc, and most stars $<$ 1 Gyr in the simulation are born within the central 1 kpc.
The right panel shows the radial migration strength ($\sigma_{\rm RM}$), calculated with 1.5*MAD $R-R_b$ of mono-age populations, using the true \rb\ and the inferred \rb.
Uncertainties in radial migration strength from the inferred \rb\ are obtained via bootstrapping, perturbing \rb\ within 1.5 kpc (the variance from Figure~\ref{fig:2}) and age within 20\% (the typical observational age uncertainty).
To account for the variance between true and inferred \rb, 1.5 kpc was subtracted from the inferred measurement. After this correction, the recovered migration strength reproduces the overall shape, including the minimum near 1 Gyr, and agrees with the true values within the estimated uncertainties.

\subsection{Recovering Birth Radii for the LMC} \label{subsec:recover_obs}
Before applying the method to observational data, we account for the selection function by first identifying the star that has the smallest represented stellar mass. 
We then make copies of the remaining stars by the number of smallest represented stellar masses they represent. 
We perturb the age and abundances within their uncertainty before adding each copy.
This increases the number of stars in the sample from $\sim$6,000 to $\sim$26,000.

\begin{figure*}[ht!]
    \centering
    \includegraphics[width=\textwidth]{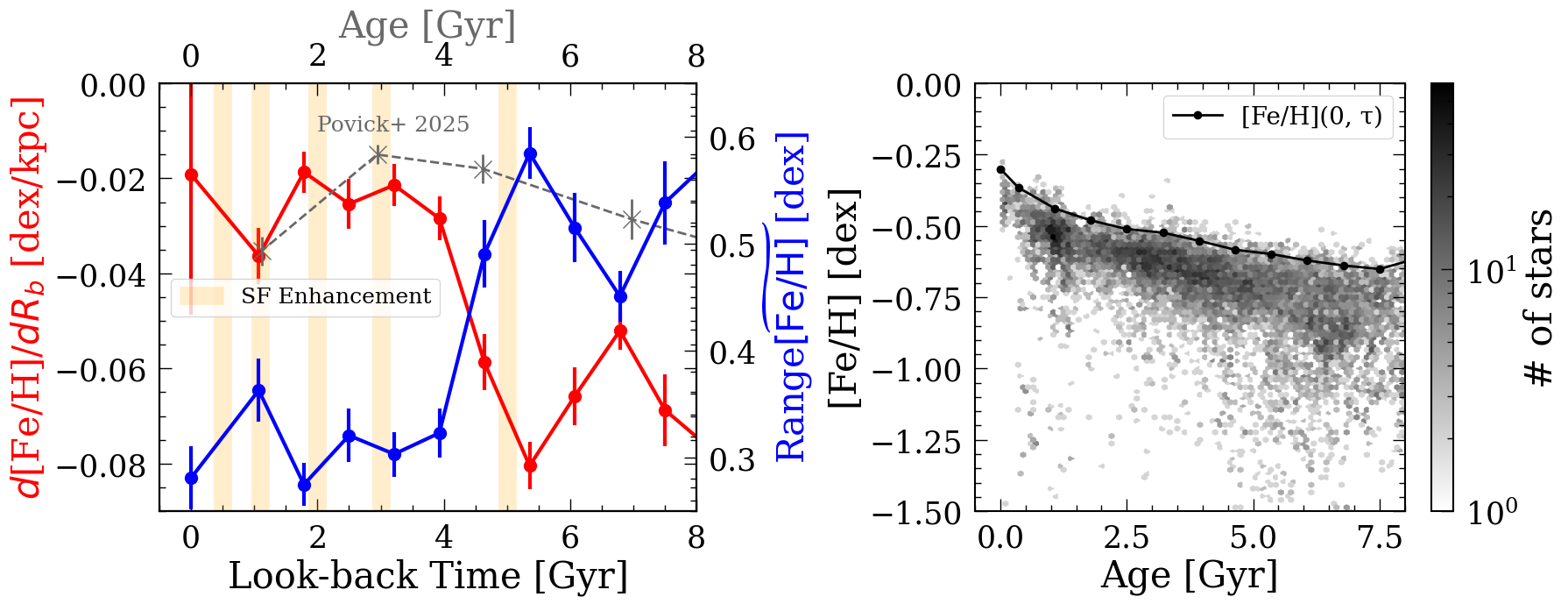}
    \caption{Similar to Figure~\ref{fig:1} but with the LMC observational data.
    The red solid line shows the inferred metallicity gradient as a function of look-back time, derived using the range in metallicity.
    The orange shaded areas are enhancements in star formation (SF) at 0.5, 1.1, 2, 3, and 5~Gyr \citep{Massana2022}.
    The gray dashed line shows the metallicity gradient as a function of age taken from \cite{Povick2023}.
    The observed metallicity gradient as a function of age is shallower than the inferred metallicity gradient at birth due to radial migration.
    Steepening in the metallicity gradient, defined as where the local minima are, is observed at or close to the times of increased SF around 1 and 5 Gyr.
    This is also seen in the MW \citep[e.g.,][]{Lu2024} and MW-like simulations \citep[e.g.,][]{Buck2023, Ratcliffe2024}.}
    \label{fig:3}
\end{figure*}

After accounting for the selection function, we apply the method described in the previous section to the data, using age bins between 0 and 8 Gyr, with a width of 0.5~Gyr to obtain Range[Fe/H].
To calculate the uncertainty, for each age bin, we performed bootstrapping 100 times by randomly selecting 200 stars each time and recomputed the percentiles.
For reference, we have over 800 stars in each age bin, and changing the number of stars we select to be 100 or 400 does not impact our result.
The uncertainty is then taken to be the standard deviation of all the calculations.
Figure~\ref{fig:3} left panel shows the measured Range[Fe/H] as a function of age (top axis). 
The center metallicity evolution is determined using the 80$^{\rm th}$-percentile of the metallicity in mono-age bins, as we did for the simulation.
The result is shown as the black solid line in the right panel of Figure~\ref{fig:3}.

\begin{figure}[ht!]
    \centering
    \includegraphics[width=\linewidth]{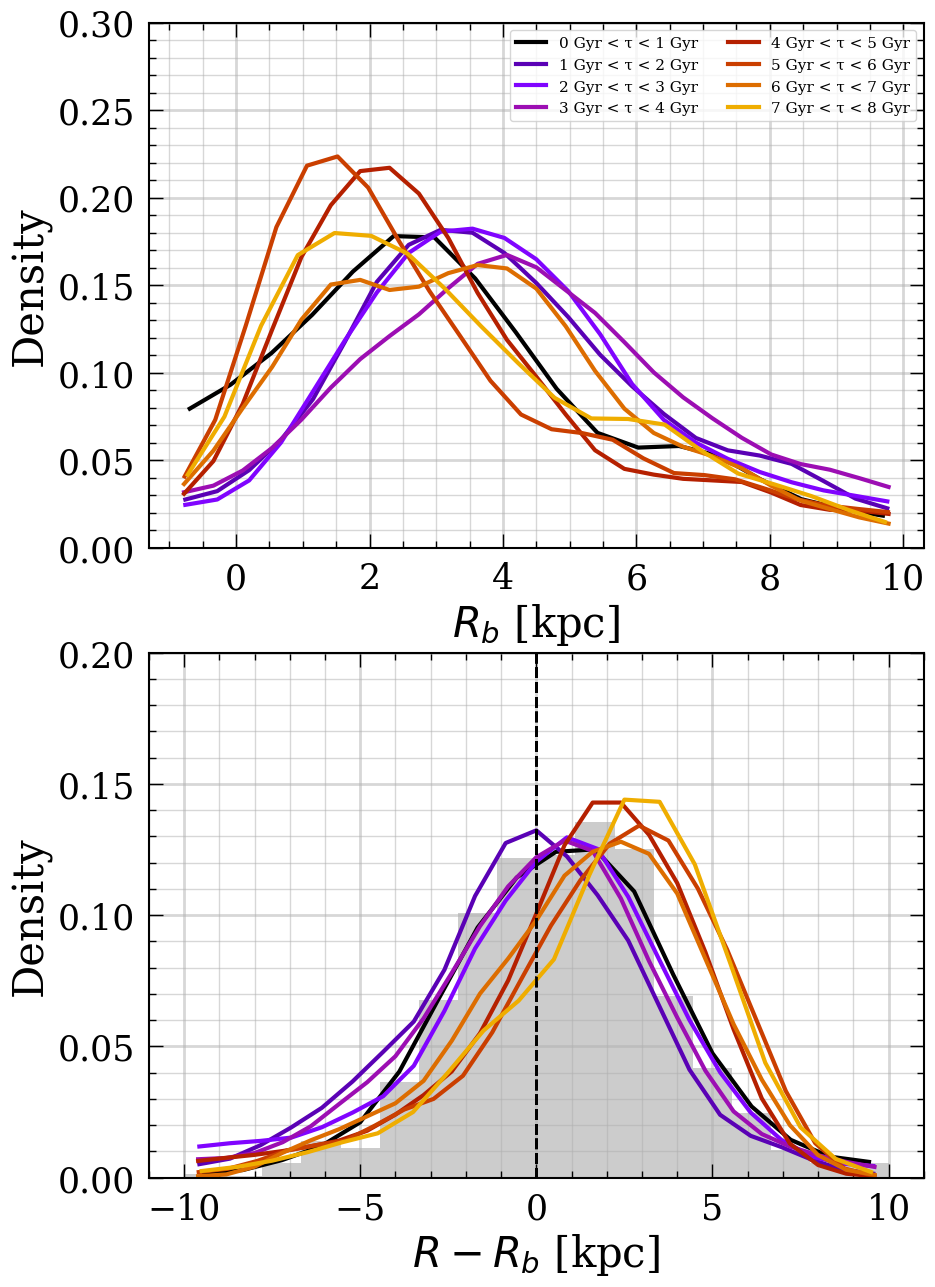}
    \caption{Normalized histograms showing the distribution of birth radii (top) and the amount of migration (bottom) for stars of different ages (as indicated in the legend).
    For each histogram, we selected the number of bins to be the integer part of $\sqrt{N}/2$ where $N$ is the number of stars in each age bin.
    We then plotted the smoothed histograms using a 1D Gaussian kernel with a width of 1 kpc for better visualization. 
    The gray histogram in the bottom panel shows the unsmoothed distribution for stars younger than 1 Gyr.
    We expect the migration distribution of these youngest stars to peak near 0 kpc with a characteristic width of $<$ 2 kpc, reflecting their limited movement since birth.}
    \label{fig:4}
\end{figure}

We inferred the birth metallicity gradient as a function of look-back time using Equation~\ref{eq1}, where $a = -0.2$ kpc$^{-1}$, and $\nabla$[Fe/H](0) = ($-0.2$ Range[Fe/H](0)$-0.022)$ dex/kpc, where $-0.022$ is the current day metallicity gradient, and Range[Fe/H](0) = $0.29$ dex is the current day Range[Fe/H],  both measured using stars $<$ 0.7 Gyr.
This ensures Equation~\ref{eq1} can reproduce the current day gradient with the current day Range[Fe/H].
We determine the value of $a$ so that the older stars are born in the inner galaxy, following inside-out growth; stars born today have not migrated much, meaning the $R-R_b$ distribution peaks close to 0 and the width of the Gaussian is $<$2 kpc for the youngest stars ($<$ 1 Gyr; see Figure~\ref{fig:4}); and the birth metallicity gradient at a look-back time is at least as steep as the observed metallicity gradient for stars of the same age, since radial migration has the effect of making the birth gradient shallower --- a natural phenomenon in galaxies with decreasing surface density towards the disk outskirt.
The uncertainty on the inferred birth metallicity gradient is determined by bootstrapping the slope 100 times with Range[Fe/H] measurements perturbed within their uncertainty. 
We mark the times (0.5, 1.1, 2, 3, and 5~Gyr) where star formation (SF) is enhanced in the LMC \citep{Massana2022}, in which the enhancements of 0.5, 1.1, 2, and 3 Gyr are thought to be caused by the interaction between the MCs, as similar peaks are also found in the SMC.
It is unclear what causes the enhancement at 5~Gyr, as this is only seen in the LMC but not in the SMC.
However, \cite{Saroon2025} finds enhancement in the SMC around the same time in the outskirts of the galaxy.

Steepening in the metallicity gradient is observed at 1 and 5 Gyr.
This is also seen in the MW \citep[e.g.,][]{Lu2024} and MW-like simulations \citep[e.g.,][]{Buck2023, Ratcliffe2024}.
They are typically caused by the sudden increase in the cold gas disk size from lower metallicity gas accreted from infalling satellites.
As a result, even if the SMC did not show enhancement in SF in its main body, the signature at 5 Gyr still matches with those of a gas accretion event.

Finally, we estimated the birth radii of stars in the LMC using Equation~\ref{eq2}.
For each star, we performed 50 Monte Carlo realizations of the birth radius calculation, drawing age, [Fe/H], and $\nabla$[Fe/H]($\tau$) from Gaussian distributions centered on their measured values and with standard deviations equal to their respective uncertainties.
The final birth radius and its uncertainty were taken as the median and half of the difference between the 84$^{\rm th}$-percentile and the 16$^{\rm th}$-percentile of these 50 realizations.
Figure~\ref{fig:4} presents the resulting birth radius and migration distributions for stars grouped by age.
As expected from inside-out galaxy formation, older stars should originate in the inner disk relative to younger stars.
Additionally, the youngest stars, which haven't had much time to migrate, exhibit a median radial migration close to 0 kpc, with a characteristic dispersion (migration strength) of approximately 1.5 kpc.

It is worth pointing out that \cite{Gallart2009}, and recently, \cite{Frankel2025}, suggested that the LMC could form outside-in based on the scale length in mono-[Fe/H], mono-age populations.
We could not test this in this work, as overall inside-out formation is one of our assumptions.
However, according to individual \rb\ distributions, the youngest stars are formed more in the inner disk, and some of the oldest stars ($>$ 6 Gyr) are formed in the outer disk.
We will revisit this topic in Section~\ref{subsec:rb_age}.

\section{Results} \label{sec:results}
\subsection{Where and When are the LMC Stars Formed}\label{subsec:rb_age}
\begin{figure*}
    \centering
    \includegraphics[width=\linewidth]{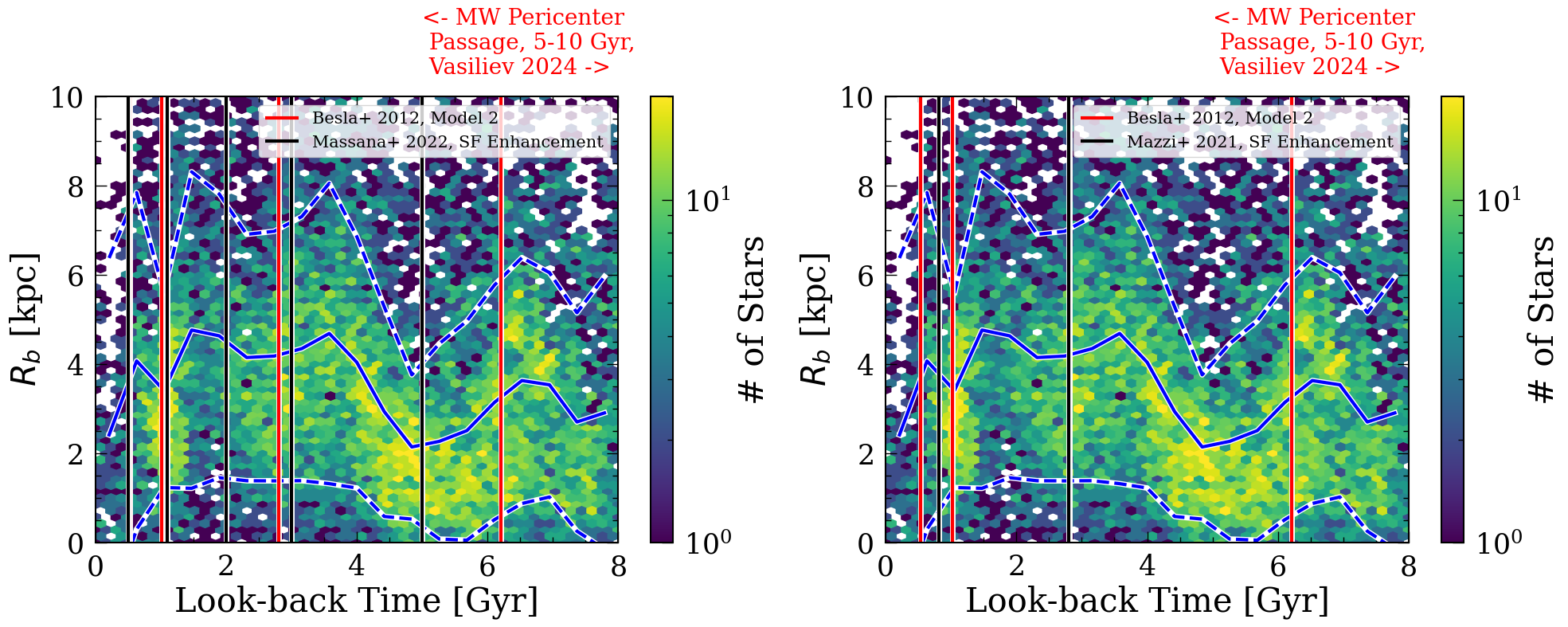}
\caption{The detailed star formation history of the LMC compared with star formation enhancements from \cite{Massana2022} and \cite{Mazzi2021}. 
The blue solid line shows the median star-forming \rb, and the dashed lines show 1.5*MAD.
The black lines are the center of the star formation episodes identified from the literature as indicated in the legends, the SMC pericenter passage times are taken from model 2 predictions in \cite{Besla2012}, and the possible first passage time with the MW 5-10 Gyr ago is shown on the top \citep{Vasiliev2024}.
The steepening in the metallicity gradient and the increase in star formation around 1.1, 3, and perhaps 5~Gyr match our expectations for an infalling satellite with varying disk alignments.
In this case, it is likely the SMC or the LMC passage with the MW.
The location of the starbursts suggests an aligned disk spin orientations between the merging galaxies during the 5 and 1.1 Gyr passages, and misaligned or perpendicular disk spin orientations for the 3 Gyr passage based on results from simulation analysis \citep{Moreno2015}.
The enhancements at 0.5 and 2 Gyr taken from \cite{Massana2022} have no significant signatures in our \rb\ distribution.
It is also interesting to point out that the star formation happens at the outer disk for the three pericenter predictions, suggesting the alignments between the LMC and SMC disks do not change if the predictions from the model are correct.}
    \label{fig:5}
\end{figure*}

With birth radii, we can now picture a more detailed star formation history --- not only when stars are formed, but also where.
Figure~\ref{fig:5} shows \rb\ as a function of look-back time, while the black vertical lines show the center of enhancement of SF episodes indicated in \cite{Massana2022} or \cite{Mazzi2021}, the red vertical lines show the model predictions of the pericenter passages of the SMC to the LMC from model 2 in \cite{Besla2012}, and the text on top shows the possible first pericenter passage of the LMC with the MW predicted in \cite{Vasiliev2024}.
The blue solid line and the blue dashed lines show the median and 1.5*MAD of \rb.
Combining information shown for the birth metallicity gradient in Figure~\ref{fig:3}, we can conclude that $>$ 6 Gyr, the LMC forms inside-out, with most of the stars forming with galactic radius $<$ 2 kpc at 8 Gyr to $<$ 6 kpc at 6 Gyr.

During the first enhancement of star formation at $\sim$5 Gyr according to \cite{Massana2022}, most stars are born in the inner disk while the metallicity gradient steepens and the center metallicity slightly increases.
Enhancement of SF in the central kpc and suppression of SF in the outer disk could happen during a passage where the primary and secondary galaxies have aligned disk spin orientations, and gas from the secondary quickly funnels in the center of the primary \citep[e.g.,][]{Moreno2015}.
If this is caused by the passage of the SMC, similar enhancement should be seen in the SMC.
This is seen in \cite{Rubele2018} but is not in \cite{Massana2022}, possibly because of the difference in spatial selection.
After this starburst, the LMC gradient then quickly flattens between 4-5 Gyr.
During the SF enhancement at $\sim$3 Gyr, most stars are formed in the center or outer disk, between 2-6 kpc.
The metallicity gradient also decreases, showing obvious expectations from gas-rich merging satellite \citep[e.g.,][]{Torrey2012, Lu2024, Buck2023}.
If this is triggered by the pericenter passage of the SMC, the disk spin orientations between the MCs are likely not aligned or perpendicular. 
The enhancement at $\sim$1.1~Gyr shows even a stronger steepening of the metallicity gradient, with stars forming across the LMC and a significant enhancement in star formation in the center, between 1-4 kpc, suggesting aligned disk spin orientations.
Finally, the enhancements identified by \cite{Massana2022} at $\sim$0.5 Gyr and 2 Gyr show no significant imprints in our \rb\ distribution.
This suggests the SMC and LMC close encounters likely happened around the enhancement in SF $\sim$1, 3 Gyr, and perhaps 5 Gyr, agreeing with predictions from some simulations \citep[e.g.,][]{Besla2012, Pardy2018}.
The star formation region also suggested a change in disk alignments between the MCs during the multiple passages. 

However, if the star formation enhancements do not track the pericenter passages of the SMC, and the model 2 predictions from \cite{Besla2012} are correct, the SMC/LMC disks are likely aligned across all passages.
This can be inferred as during all three of these pericenter passages, most stars formed in the middle or outer disk, suggesting disk alignments.
The size of the star-forming region (indicated by the blue dashed lines) also increased close to the times when the SMC was predicted to be at the pericenter passages with the LMC according to model 2.

\subsection{Radial Migration Strength in the LMC}\label{subsec:rb_age}
\begin{figure*}
    \centering
    \includegraphics[width=0.55\linewidth]{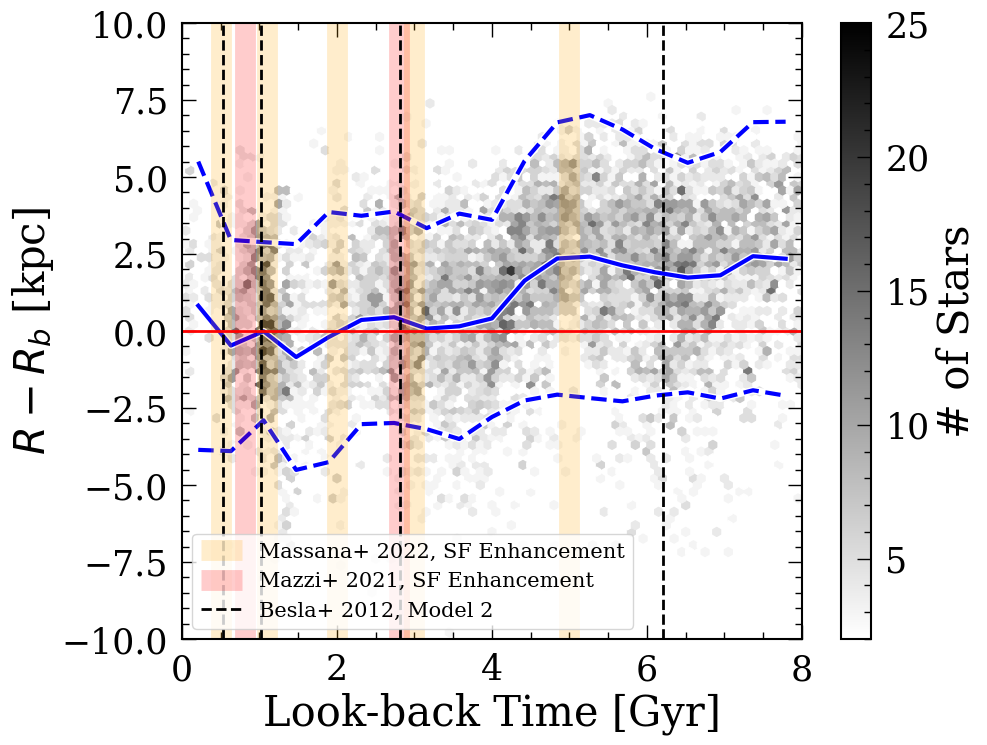}
    \includegraphics[width=0.42\linewidth]{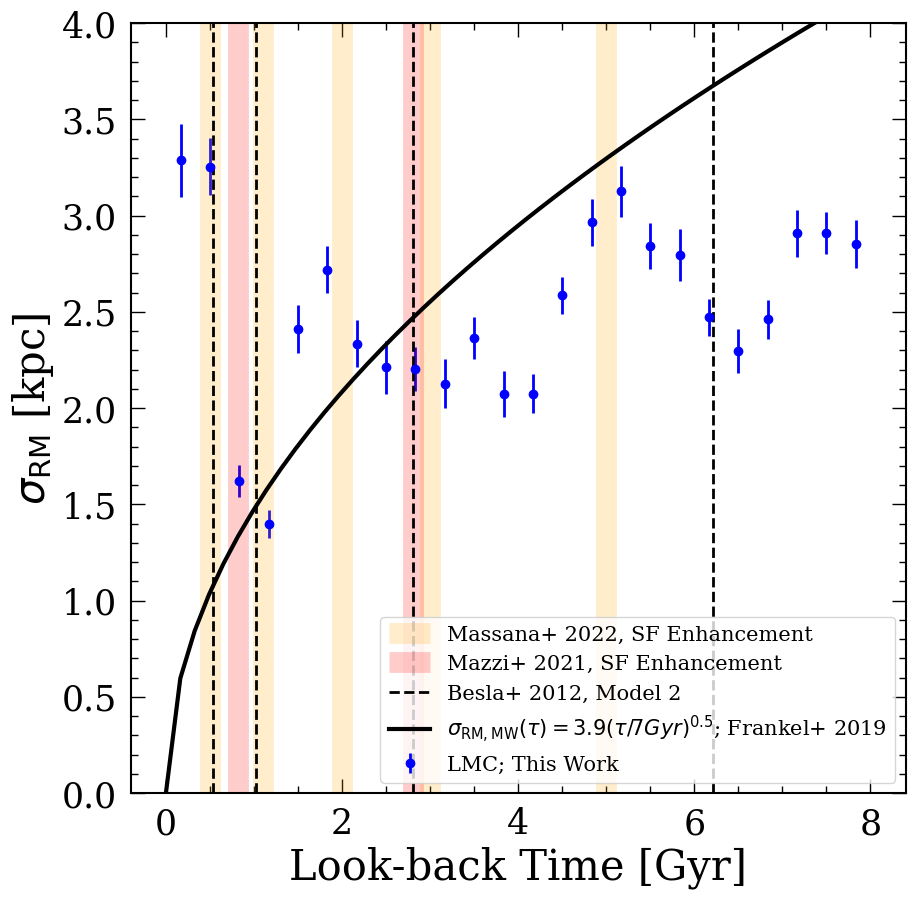}
\caption{
Left: $R-R_b$ for the LMC stars, where the blue solid line shows the median and the blue dashed line shows the 1.5*MAD. The orange and red vertical lines show the SF enhancements taken from \cite{Massana2022} and \cite{Mazzi2021}, respectively.
The black dashed lines show the model prediction from \cite{Besla2012}.
Right: Radial migration strength as a function of look-back time for the LMC data (blue points).
We subtracted 1.5 kpc from the migration strength for the LMC data, assuming we are able to infer \rb\ to a similar level of uncertainty as the simulation.
Note, this shift will only affect the constant in front of the power-law fit, but not the power index.
The black line shows the analytic solution of the radial migration strength for the MW, taken from \cite{Frankel2019}.
The LMC stars show an increase in migration strength around 0.5, 2, and 5 Gyr, close to the SF enhancements around the same time.
The power index for the migration strength exhibits a smaller value compared to the MW.
}
    \label{fig:6}
\end{figure*}

We quantify the radial migration strength by looking at the median and width of the $R-R_b$ distributions as a function of look-back time, where the median represents the overall migration of a mono-age population, and the width quantifies the strength. 
Figure~\ref{fig:6} shows the radial migration for the LMC stars as a function of look-back time, where the left plot shows the full distribution with median (blue solid line) and 1.5*MAD (blue dashed line), and the right plot shows the standard deviation ($\sigma_{\rm RM}$) as a function of look-back time for the LMC data (blue points).
For comparison, we also plotted the analytic model for the MW taken from \cite{Frankel2019}.
The standard deviation is measured to be 1.5*MAD to account for outliers.
The orange and red lines are the SF enhancements taken from \cite{Massana2022} and \cite{Mazzi2021}, respectively.
We also plotted the predictions for the pericenter passages taken from \cite{Besla2012} model 2.

We find an increase in radial migration strength close to the SF enhancements at 0.5, 1.1, and 5 Gyr.
During these times, stars migrated outwards, exceeding the average.
However, during the enhancements around 2 and 3 Gyr, the stars have median migrations close to 0.
This is not surprising as most stars in our sample are born in the inner disk during the increase in SF around 1.1, and 5 Gyr (see Figure~\ref{fig:5}), meaning they can only migrate outwards.
Since stars are born more evenly across the disk at 3 Gyr, the median stays around 0.
The right panel of Figure~\ref{fig:6} suggests the migration strength is significantly higher close to most SF enhancements.
This is especially obvious at 0.5, 2, and 5 Gyr.
Interestingly enough, 0.5 and 2 Gyr are the times that we find no significant preference for where in the disk the stars are preferably born.
These increases in migration strength can be caused by external perturbation of the SMC on the LMC during pericenter passages \citep[e.g.,][]{Quillen2009}.

The radial migration strength as a function of time for the LMC could be similar, but with a slightly shallower slope than the MW analytic solution taken from \cite{Frankel2019}.
The migration strength is consistently higher compared to the migration strength from the MW data and analytic solution during the time of frequent SF enrichment below $\sim$~1~Gyr, where the MCs are most likely interacting.
This suggests interacting galaxies of mass ratio 1:10 can produce a significant increase in migration strength. 

\subsection{The Chemodynamic History of the LMC}
With \rb, we are able to recover the Chemodynamic history of the LMC empirically.
We do so by looking at the age-metallicity, age-\alpham, and [Fe/H]-\alpham\ relations in mono-\rb\ and mono-age populations. 

\begin{figure*}
    \centering
    \includegraphics[width=0.48\linewidth]{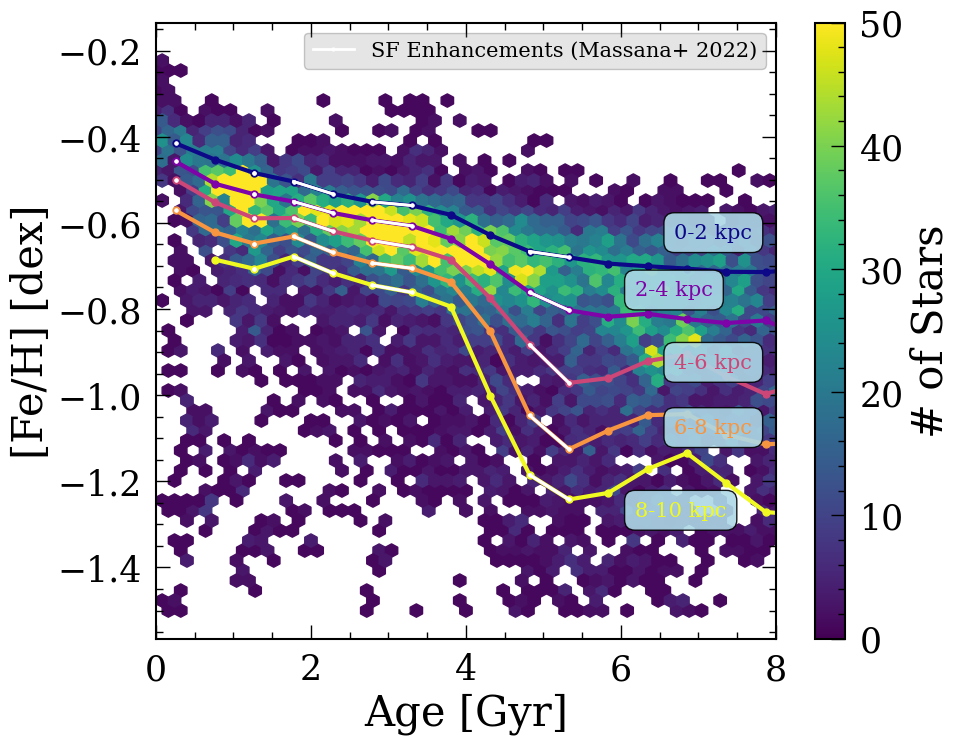}
    \includegraphics[width=0.48\textwidth]{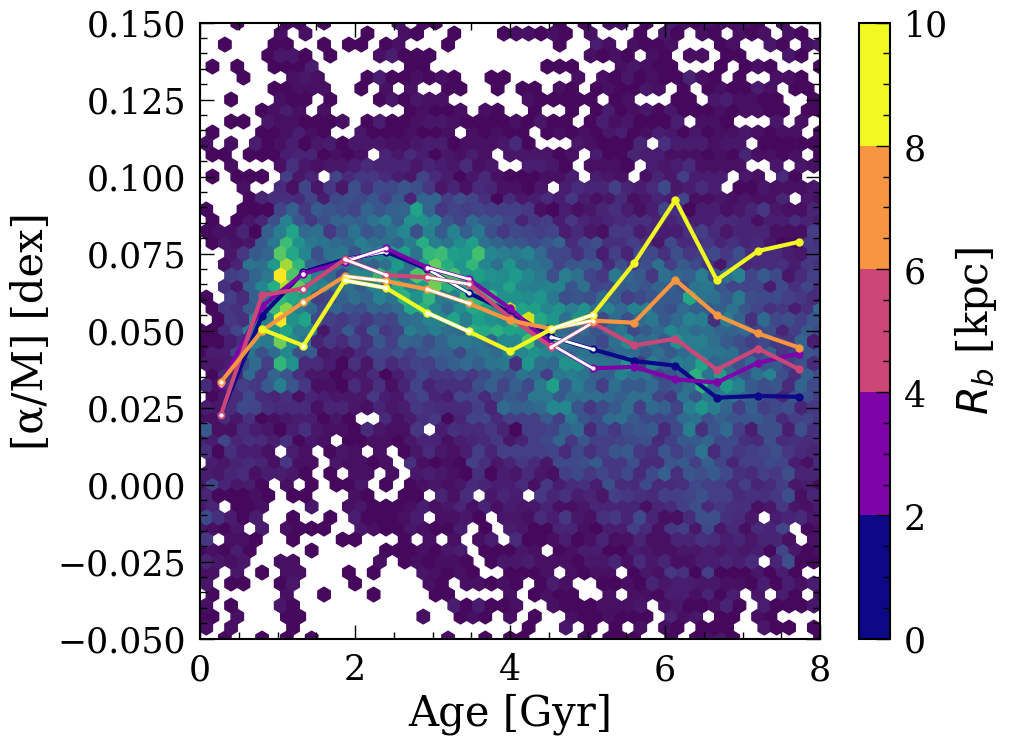}
    \caption{The age-metallicity (left) and the age-\alpham\ relations plotted as density in the background, and the colored lines show the median temporal evolution in each mono-\rb\ population. 
    The points on each line are separated equally in age with a difference of 0.5 Gyr.
    These tracks place additional constraints on the gas accretion and star formation history at various locations and times.
    For example, combining with the star formation region shown in Figure~\ref{fig:5}, at 5 Gyr, the LMC accreted a large amount of pristine gas at all radii (drop in [Fe/H] in all radii in the left panel) that increased the star formation efficiency in the inner disk (from Figure~\ref{fig:5}). 
    This led to a decrease in \alpham\ in the outer disk since there was no star formation, and an increase of \alpham\ in the inner disk where most stars were formed, contributing to the $\alpha$ element reservoir via SN Type II.}
    \label{fig:7}
\end{figure*}

Figure~\ref{fig:7} shows the age-metallicity (left) and age-\alpham\ (right) relations in mono-\rb\ and mono-age populations.
Each colored line shows the median evolution of the abundances of stars in a 2 kpc \rb\ bin.
The white segments show the times at which SF enhancements are observed.
During the enhancements at 1, 3, and 5 Gyr, the overall metallicity goes down or stays constant at all radii, suggesting dilution from pristine gas. 
This also matches with the enhancement of SF during those times, as observed in Figure~\ref{fig:5}.
Since the majority of star formation 5 Gyr ago is concentrated in the inner disk, $\alpha$ elements should be produced mostly in the central region.
As a result, \alpham\ only increases for mono-\rb\ tracks $<\sim$~5~kpc, and \alpham\ decreases in the outer disk, where star formation efficiency (SFE) is low, and no $\alpha$ elements are produced. 
At 3 Gyr, the star formation spreads across the entire disk, thus, \alpham\ increases for almost all mono-\rb\ tracks, thanks to the contribution of SN Type II from these newly formed stars.
At 1 Gyr, it is possible that the LMC did not accrete a significant amount of gas, given the minor decrease in [Fe/H].
The decrease of \alpham\ at all radii during that time suggests the SFE is not high.
These tracks place strong constraints on the gas accretion and star formation history at various locations and times.

Figure~\ref{fig:8} shows the temporal evolution of \alpham-[Fe/H] for various radii.
The biggest panel on the top right shows the mono-\rb\ tracks, similar to what is done in Figure~\ref{fig:7}.
The smaller panels show the density and median abundances for stars in different \rb\ bins, increasing from the top left to the bottom right.
The white lines and points mark the SF enhancements.
Similar to what was found in the MW in \cite{Lu2024}, the SF enhancements naturally show up as over-densities in mono-\rb\ populations.
The most $\alpha$-enriched stars are formed during the SF enhancements 3-2~Gyr ago, mostly between 2-4~kpc.
This is also the time when most of the stars are formed during the SF enhancement at 3~Gyr.

\section{Insight into the lack of \alpham-[Fe/H] bimodality of the LMC}\label{sec:twoinfall}
The MW exhibits a distinct gap between the high- and low-$\alpha$ disk \citep[e.g.,][]{Bensby2014, Nidever2014, Hayden2015}, and popular theories include the two-infall model \citep[e.g.,][]{Chiappini1997, Spitoni2021}, clumpy star formation \citep[e.g.,][]{Clarke2019, Garver2023}, radial migration \citep[e.g.,][]{Sharma2020}, and quenching in star formation \citep[e.g.,][]{Beane2025}.
Compared to the MW, the LMC does not show such a signature, which raises questions on whether the bimodality is a common feature in disk galaxies \citep[see also analysis on M31 in][]{Nidever2024}.

\begin{figure*}
    \centering
    \includegraphics[width=0.95\textwidth]{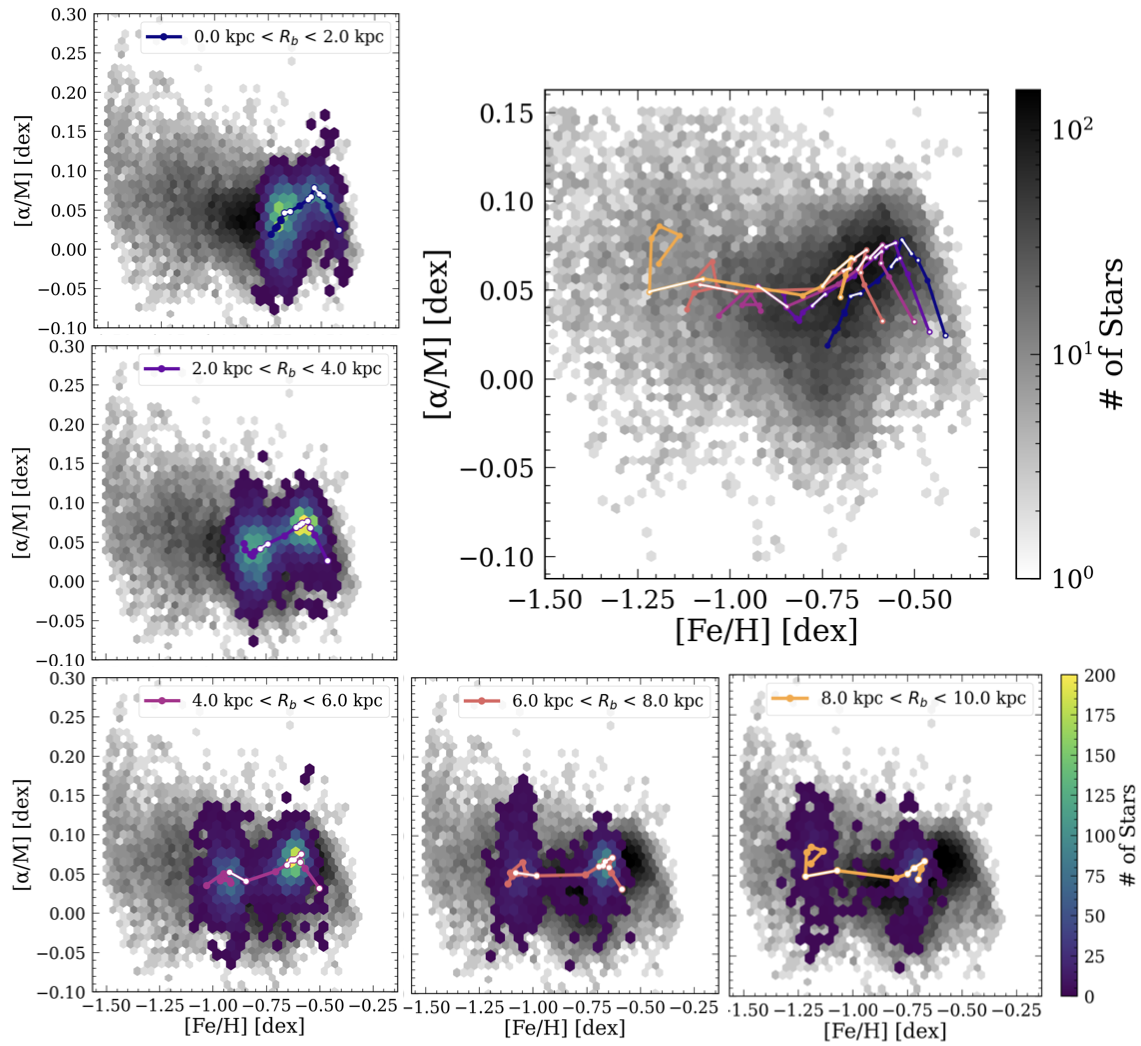}
    \caption{
    The largest panel on the top right shows the median mono-\rb\ tracks in \alpham-[Fe/H] space, similar to what is done in Figure~\ref{fig:7}.
    The smaller panels show the density and median abundances for stars in different \rb\ bins, increasing from the top left (inner galaxy) to the bottom right (outer galaxy).
    The white lines and points mark the SF enhancements taken from \cite{Massana2022}.
    }
    \label{fig:8}
\end{figure*}

What is interesting is that looking at LMC stars in mono-\rb\ population in Figure~\ref{fig:8}, overdensities in [$\alpha$/M]-[Fe/H] show up during times of SF enhancements, similar to what is seen in the MW \citep{Lu2024}.
We compare the most prominent steepening of the metallicity gradient in both galaxies --- at 5 Gyr ago for the LMC and 8 Gyr ago for the MW --- during which the MW developed its low-$\alpha$ disk.
Examining Figure 7 in \citet{Lu2024}, we suggest that the distinct bimodality observed in the MW's \alpham–[Fe/H] plane, compared to the LMC, may be attributed to the extended star formation history in the MW's outskirts.
This extended star formation began around 8 Gyr ago and directly produced the most metal-poor low-$\alpha$ stars.
In contrast, star formation in the outskirts of the LMC appears to have been suppressed during its corresponding gradient-steepening episode at 5 Gyr, preventing the formation of a similarly extended low-$\alpha$ sequence.
Moreover, the centrally concentrated star formation in the LMC at this time mitigated the typical decline in \alpham\ caused by delayed SN Ia enrichment.
The simultaneous increase in $\alpha$-element production from SN II explosions associated with new star formation counteracts the SN Ia effect, resulting in little net change in \alpham\ in the inner disk during the 5 Gyr event, preventing any bimodality from forming in the inner disk. 

In summary, the lack of bimodality in the \alpham-[Fe/H] plane for the LMC could likely be caused by a combination of the changes in star-forming regions and enrichment history.

\section{Limitations}
The main caveat is whether we can robustly recover \rb\ for the LMC, given its recent interaction with the SMC.
Tests on the NIHAO simulation suite \citep{Wang2015} suggest this should be feasible. In \cite{Lu2024_lmc}, we showed that an LMC-like galaxy—with an estimated dark matter halo mass of $18.8^{+3.5}{-4.0}\times10^{10},M\odot$ \citep[][]{Shipp2021} and a current “diskiness” of $\langle v_\phi \rangle/\sigma_v \sim 2.8$ \citep[Figure B.5 in][]{gaiaLMC2021}—can be used to infer \rb\ using the method described in \cite{Lu2024}.
In that work, we successfully recovered \rb\ for a simulated LMC-mass galaxy (halo mass $1.64\times10^{11}, M_\odot$) that had undergone a merger roughly 1 Gyr ago, similar to the recent interaction between the Magellanic Clouds.
Furthermore, the simulation analyzed in this study \citep{Garver2026}, which reproduces the interaction history of the MCs, also indicates that we should be able to robustly recover \rb\ for the LMC and extract meaningful insights into its migration and star formation history.
That said, our ability to validate the method is limited to simulations, and unforeseen effects from the MCs’ interaction could hinder its applicability.
Additional assumptions, such as the metallicity gradient being negative and approximately linear during the early formation of the LMC’s stellar disk, may also introduce uncertainties.

\section{Conclusion \& Future Work}
In this paper, we tested a previous method described in \cite{Lu2024} to infer birth radii (\rb) for star particles in a hydrodynamical simulation intended to reproduce the interaction history of the MCs.
We then applied this method to infer birth radii for the LMC stars.
Combining \rb\ and age information, we are able to study the chemodynamic history of the LMC in the context of its interaction with the SMC.
We found:
\begin{itemize}
    \item Steepening of the LMC metallicity gradient around 5, 3, and 1~Gyr ago, suggesting accretion of pristine gas. 
    These times coincide with some of the SF enhancements reported by \cite{Massana2022}.
    \item At 5~Gyr, star formation is concentrated in the inner disk. 
    By 3~Gyr, it becomes more widespread, extending from the center to the outer disk. 
    At 1~Gyr, the star formation remains broadly distributed but shows renewed concentration in the inner disk.
    These differences, if caused by the interactions between the MCs, could be due to different alignments of the disk spin orientations \citep[e.g.,][]{Moreno2015}.
    However, if the SF enhancements do not track the pericenter passages of the SMC to the LMC, and the inferred passages are correct in model 2 from \cite{Besla2012}, the SMC and LMC disk spin orientations are likely aligned and has not changed, as the star formations are concentrated in the middle and outer disk for all three predicted pericenter passage times.
    \item The migration strength for the LMC is found to be 4.4($\tau$/8 Gyr)$^{0.4}$, with significant enhancements around 0.5, 2, and 5~Gyr.
    \item No strong signature of the SF enhancements is found in the metallicity gradient or preferred star formation location at 0.5 and 2~Gyr, but an increase in migration strength is observed for both.
    \item SF enhancements show up as clumps in \alpham-[Fe/H] space, as does for the MW in \cite{Lu2024}.
    \item The most $\alpha$-enriched stars are formed 2-3~Gyr ago, mostly between 2-4~kpc.
    This is also the only time when the star formation is spread out across the disk.
    \item The LMC does not exhibit a clear \alpham–[Fe/H] bimodality like that observed in the Milky Way, despite showing similar signatures of significant gas accretion events.
    This difference is likely due to the centrally concentrated nature of star formation in the LMC during these episodes, in contrast to the more spatially extended star formation that occurred in the outskirts of the MW.
\end{itemize}

These relations place strong constraints on the formation of the LMC and the interaction history between the MCs. 
Future work should explore different simulation suites to better understand how merger orbits and gas accretion events (e.g., galactic fountains) can affect the star formation region in the merging galaxies and explore different assumptions, as in this work, we assume stars preferentially migrate outwards as a population.

\begin{acknowledgments}
\section{Acknowledgments}
Y.L. wants to thank James Johnson for the helpful discussion. 
B.G. and D.L.N. acknowledge support by NSF grants AST 1908331 and 2408159.
We thank V. Debattista for help running the Gasoline simulations.
NGC acknowledges support provided by the Heising-Simons Foundation grant \# 2022-3927.
This work has made use of data from the European Space Agency (ESA) mission Gaia,\footnote{\url{https://www.cosmos.esa.int/gaia}} processed by the Gaia Data Processing and Analysis Consortium (DPAC).\footnote{\url{https://www.cosmos.esa.int/web/gaia/dpac/consortium}} 
Funding for the DPAC has been provided by national institutions, in particular the institutions participating in the Gaia Multilateral Agreement.
This research also made use of public auxiliary data provided by ESA/Gaia/DPAC/CU5 and prepared by Carine Babusiaux. 
This research has also made use of NASA's Astrophysics Data System.

\end{acknowledgments}

%

\vspace{5mm}
\facilities{Gaia \citep{gaia, Gaiadr3}, 2MASS \citep{2mass}, APOGEE--2S \citep{Wilson2019}}


\software{astropy \citep{astropy:2013, astropy:2018, astropy2022}, Matplotlib \citep{matplotlib}, NumPy \citep{Numpy}, Pandas \citep{pandas}, \texttt{ChatGPT} \citep{openai2025chatgpt}
          }






\bibliography{sample631}{}
\bibliographystyle{aasjournal}



\end{document}